\documentclass[twocolumn,notitlepage,nofootinbib]{revtex4-2}
\usepackage{amsmath,amssymb,bm,graphicx,hyperref,slashed}
\allowdisplaybreaks[1]

\renewcommand\d{\partial}
\newcommand\grad{\bm{\nabla}}

\newcommand\0{\bm{0}}
\newcommand\A{\bm{A}}
\newcommand\B{\bm{B}}
\renewcommand\j{\bm{j}}
\renewcommand\k{{\bm{k}}}
\newcommand\p{{\bm{p}}}
\newcommand\q{{\bm{q}}}
\newcommand\x{\bm{x}}
\newcommand\Z{\mathbb{Z}}
\newcommand\C{\mathcal{C}}
\renewcommand\P{\mathcal{P}}
\newcommand\D{\mathrm{D}}
\newcommand\PV{\mathrm{PV}}
\newcommand\reg{\mathrm{reg}}
\DeclareMathOperator\arctanh{arctanh}
\DeclareMathOperator\diag{diag}
\DeclareMathOperator\tr{tr}

\begin{document}

\title{Torsion-induced chiral magnetic current in equilibrium}

\author{Tatsuya Amitani}
\author{Yusuke Nishida}
\affiliation{Department of Physics, Tokyo Institute of Technology,
Ookayama, Meguro, Tokyo 152-8551, Japan}

\date{April 2022}

\begin{abstract}
We study equilibrium transport properties of massless Dirac fermions at finite temperature and chemical potential in spacetime accompanied by torsion, which in four dimensions couples with Dirac fermions as an axial gauge field.
In particular, we compute the current density at the linear order in the torsion as well as in an external magnetic field with the Pauli-Villars regularization, finding that an equilibrium current akin to the chiral magnetic current is locally induced.
Such torsion can be realized in condensed matter systems along a screw dislocation line, around which localized and extended current distributions are predicted so as to be relevant to Dirac and Weyl semimetals.
Furthermore, we compute the current density at the linear order in the torsion as well as in a Weyl node separation, which turns out to vanish in spite of being allowed from the symmetry perspective.
Contrasts of our findings with torsion-induced currents from previous work are also discussed.
\end{abstract}

\maketitle
\tableofcontents

\section{Introduction}
Massless Dirac fermions exhibit both the vector and axial U(1) symmetries at a classical level.
However, the latter is violated by a quantum effect known as the axial anomaly~\cite{Adler:1969,Bell:1969}.
The axial anomaly is not only important in particle physics~\cite{Witten:1979,Veneziano:1979}, but also serves as a source of anomalous transport phenomena in relativistic quantum matters~\cite{Vilenkin:1980,Nielsen:1983}.
One celebrated example is the so-called chiral magnetic effect~\cite{Kharzeev:2008,Fukushima:2008}, which predicts the electric current along an external magnetic field in the presence of an axial chemical potential:
\begin{align}\label{eq:CME}
j^i_\mathrm{CME} = \frac{\mu_5}{2\pi^2}B^i.
\end{align}
Up to now, the chiral magnetic effect has attracted broad interest because it is potentially relevant to diverge systems~\cite{Miransky:2015}, such as quark-gluon plasmas, neutron stars, and even Dirac and Weyl semimetals realizing massless relativistic fermions in condensed matter systems~\cite{Armitage:2018}.

The chiral magnetic current in Eq.~(\ref{eq:CME}) is nondissipative if it exists in equilibrium.
However, after resolving some confusion, it is now established that the chiral magnetic effect is absent in equilibrium~\cite{Vazifeh:2013,Basar:2014,Landsteiner:2016}.
Therefore, in order to activate the chiral magnetic effect, one needs to drive the system out of equilibrium.
For example, by applying an external electric field in Dirac and Weyl semimetals~\cite{Son:2013,Burkov:2014}, possible signatures of the chiral magnetic effect have been experimentally observed~\cite{Armitage:2018}.
It was also proposed to induce the chiral magnetic effect transiently by distorting a crystal with strain~\cite{Cortijo:2016,Pikulin:2016} or by heating Weyl nodes unequally under a strong electric field~\cite{Nandy:2020}.

What we explore in this work is an equilibrium current akin to the chiral magnetic current.
In particular, we consider massless Dirac fermions in spacetime accompanied by torsion.
Although the torsion is absent in Einstein's theory of general relativity, it can be realized in condensed matter systems by dislocations of a crystal~\cite{Katanaev:2005,Kleinert}.
Because the torsion in four dimensions is known to couple with Dirac fermions as an axial gauge field~\cite{Shapiro:2002}, its temporal component serves as an axial chemical potential but with spatial dependence, which under an external magnetic field may locally induce the chiral magnetic current in equilibrium.

To this end, we proceed as follows.
After introducing Dirac fermions in curved spacetime with torsion in Sec.~\ref{sec:torsion}, the current density in the presence of the torsion as well as an external magnetic field is evaluated with the Matsubara formalism at finite temperature and chemical potential in Sec.~\ref{sec:magnetic}.
Furthermore, we explore another possible form of equilibrium current in Sec.~\ref{sec:weyl}, which is the current density in the presence of the torsion as well as a Weyl node separation.
Finally, our findings are summarized in Sec.~\ref{sec:summary}, where their contrasts with torsion-induced currents from previous work are also discussed.

We set $\hbar=c=e=k_B=1$ throughout this paper and the Minkowski metric is chosen to be $\eta_{\alpha\beta}=\diag(1,-1,-1,-1)$.
Greek indices such as $\alpha,\beta,\dots$ and $\mu,\nu,\dots$ are valued at $0,1,2,3$, whereas Latin $a,b,\dots$ and $i,j,\dots$ are at $1,2,3$, and repeated indices are implicitly summed.
An integration over three-dimensional momentum $\p$ is denoted by $\int_\p\equiv\int d^3\p/(2\pi)^3$ for the sake of brevity.

\section{Dirac fermions with torsion}\label{sec:torsion}
Let us consider massless Dirac fermions in curved spacetime~\cite{Nieh:1982,Parker-Toms}, whose hermitian action is provided by
\begin{align}\label{eq:action_old}
S &= \frac12\int d^4x\,e(x)
\bigl[\bar\psi(x)\gamma^\alpha e_\alpha{}^\mu(x)iD_\mu\psi(x) \notag\\
&\quad - iD_\mu\bar\psi(x)\gamma^\alpha e_\alpha{}^\mu(x)\psi(x)\bigr].
\end{align}
Here, $e(x)=\det[e^\alpha{}_\mu(x)]$ is the determinant of vierbein,
\begin{align}
D_\mu\psi(x) &= [\d_\mu + iA_\mu(x) + iA_{5\mu}(x)\gamma^5 - i\omega_\mu(x)]\psi(x), \\
D_\mu\bar\psi(x) &= \bar\psi(x)
[\overleftarrow\d_{\!\!\mu} - iA_\mu(x) + iA_{5\mu}(x)\gamma^5 + i\omega_\mu(x)]
\end{align}
are the covariant derivatives, both the vector and axial gauge fields are introduced for the sake of generality, and $\omega_\mu(x)=\sigma_{\alpha\beta}\omega^{\alpha\beta}{}_\mu(x)$ is the spin connection with $\gamma^5=i\gamma^0\gamma^1\gamma^2\gamma^3$ and $\sigma_{\alpha\beta}=i[\gamma_\alpha,\gamma_\beta]/8$.%
\footnote{Although the spin connection may not be realized in condensed matter systems, it is convenient to keep it temporarily to reveal the symmetry constraint discussed in the paragraph of Eq.~(\ref{eq:constraint}).}
This action is invariant under the local Lorentz transformation,
\begin{align}
\psi(x) \to L_\Lambda(x)\psi(x), \qquad
\bar\psi(x) &\to \bar\psi(x)L_\Lambda^{-1}(x),
\end{align}
provided that the vierbein and the spin connection transform as $e^\alpha{}_\mu(x)\to\Lambda^\alpha{}_\beta(x)e^\beta{}_\mu(x)$ and
\begin{align}\label{eq:lorentz}
\omega_\mu(x) \to L_\Lambda(x)\omega_\mu(x)L_\Lambda^{-1}(x)
- i[\d_\mu L_\Lambda(x)]L_\Lambda^{-1}(x),
\end{align}
respectively, where $L_\Lambda$ is the spinor representation of $\Lambda$ satisfying $L_\Lambda^{-1}(x)\gamma^\alpha L_\Lambda(x)=\Lambda^\alpha{}_\beta(x)\gamma^\beta$.
Accordingly, each component of the spin connection transforms as
\begin{align}
\omega^{\alpha\beta}{}_\mu(x)
\to \Lambda^\alpha{}_{\gamma}(x)\Lambda^\beta{}_{\delta}(x)\omega^{\gamma\delta}{}_\mu(x)
- [\d_\mu\Lambda^\alpha{}_\gamma(x)]\Lambda^{\beta\gamma}(x).
\end{align}

In terms of the vierbein and the spin connection, the metric tensor and the affine connection are expressed by $g_{\mu\nu}(x)=\eta_{\alpha\beta}\,e^\alpha{}_\mu(x)e^\beta{}_\nu(x)$ and
\begin{align}\label{eq:affine}
\Gamma^\lambda{}_{\mu\nu}(x) = e_\alpha{}^\lambda(x)
[\d_\mu e^\alpha{}_\nu(x) + \omega^{\alpha\beta}{}_\mu(x)e_{\beta\nu}(x)],
\end{align}
respectively, and its antisymmetric part defines the torsion tensor via
\begin{align}\label{eq:torsion}
T^\lambda{}_{\mu\nu}(x) = \Gamma^\lambda{}_{\mu\nu}(x) - \Gamma^\lambda{}_{\nu\mu}(x).
\end{align}
Because the affine connection in Eq.~(\ref{eq:affine}) satisfies the metric compatibility condition, $\nabla_\lambda g_{\mu\nu}=0$, it can be decomposed uniquely into the torsionless and torsionful parts as
$\Gamma^\lambda{}_{\mu\nu}(x)=\mathring\Gamma^\lambda{}_{\mu\nu}(x)+K^\lambda{}_{\mu\nu}(x)$, where
\begin{align}\label{eq:levi-civita}
\mathring\Gamma^\lambda{}_{\mu\nu}(x)
= \frac12g^{\lambda\kappa}(x)[\d_\mu g_{\nu\kappa}(x) + \d_\nu g_{\mu\kappa}(x) - \d_\kappa g_{\mu\nu}(x)]
\end{align}
is the Levi-Civita connection and
\begin{align}
K^\lambda{}_{\mu\nu}(x)
= \frac12[T^\lambda{}_{\mu\nu}(x) - T_{\mu\nu}{}^\lambda(x) - T_{\nu\mu}{}^\lambda(x)]
\end{align}
is the contorsion tensor~\cite{Nakahara}.

In order to expose the torsion hidden in Eq.~(\ref{eq:action_old}), we follow the procedure of Ref.~\cite{Shapiro:2002} and perform the integration by parts, leading to
\begin{align}
S = \int d^4x\,e(x)\bar\psi(x)\gamma^\alpha e_\alpha{}^\mu(x)
\biggl[iD_\mu + \frac{i}{2}T^\nu{}_{\mu\nu}(x)\biggr]\psi(x).
\end{align}
Then, the spin connection expressed in terms of the vierbein and the affine connection via Eq.~(\ref{eq:affine}) is decomposed as $\omega^{\alpha\beta}{}_\mu(x)=\mathring\omega^{\alpha\beta}{}_\mu(x)+e^\alpha{}_\lambda(x)K^\lambda{}_{\mu\nu}(x)e^{\beta\nu}(x)$, where the torsionless part is provided by
\begin{align}\label{eq:torsionless}
\mathring\omega^{\alpha\beta}{}_\mu(x) = e^\alpha{}_\lambda(x)
[\d_\mu e^{\beta\lambda}(x) + \mathring\Gamma^\lambda{}_{\mu\nu}(x)e^{\beta\nu}(x)].
\end{align}
Accordingly, by separating out the torsionful part from the covariant derivative, we obtain
\begin{align}\label{eq:action_new}
S = \int d^4x\,e(x)\bar\psi(x)\gamma^\alpha e_\alpha{}^\mu(x)
\bigl[i\mathring{D}_\mu - S_\mu(x)\gamma^5\bigr]\psi(x),
\end{align}
where $\mathring{D}_\mu=\d_\mu+iA_\mu(x)+iA_{5\mu}(x)\gamma^5-i\mathring\omega_\mu(x)$ and
\begin{align}\label{eq:axial}
S^\kappa(x) \equiv \frac18\epsilon^{\kappa\lambda\mu\nu}(x)T_{\lambda\mu\nu}(x)
\end{align}
are introduced with $\epsilon^{\kappa\lambda\mu\nu}(x)$ being the totally antisymmetric tensor for $\epsilon^{0123}(x)=e^{-1}(x)$.
The resulting action is again invariant under the local Lorentz transformation because $\mathring\omega_\mu(x)=\sigma_{\alpha\beta}\mathring\omega^{\alpha\beta}{}_\mu(x)$ transforms in the same way as Eq.~(\ref{eq:lorentz}).

We now find that the torsion defined in Eq.~(\ref{eq:torsion}) couples with Dirac fermions only as an additional axial gauge field via $S^\mu(x)$~\cite{Shapiro:2002}.
It is important to emphasize that the torsion-like structure such as $\d_\mu e^\alpha{}_\nu(x)-(\mu\leftrightarrow\nu)$ cannot arise in physical observables from the vierbein explicit in Eq.~(\ref{eq:action_new}).
This is because the derivative acting on such a vierbein must be accompanied by the torsionless spin connection so as to be covariant under the local Lorentz transformation,
\begin{align}\label{eq:constraint}
[\d_\mu e^\alpha{}_\nu(x) + \mathring\omega^{\alpha\beta}{}_\mu(x)e_{\beta\nu}(x)]
- (\mu\leftrightarrow\nu),
\end{align}
which however vanishes under Eqs.~(\ref{eq:levi-civita}) and (\ref{eq:torsionless}).
Therefore, as far as our interest is focused on the current density, $j^\mu(x)=\bar\psi(x)\gamma^\alpha e_\alpha{}^\mu(x)\psi(x)$, induced by the torsion but not by the Riemann curvature, it is allowed to set $e^\alpha{}_\mu(x)=\delta^\alpha{}_\mu$ for the vierbein other than in $S^\mu(x)$ so as to reduce the spacetime to being flat with the torsion fully kept as $S^\mu(x)$.
This is the simplification employed in our following analyses in Secs.~\ref{sec:magnetic} and \ref{sec:weyl}.

Finally, we show how the torsion is realized in condensed matter systems according to Refs.~\cite{Katanaev:2005,Kleinert}.
Dislocations of a crystal regarded as a continuous medium are described by the Burgers vector,
\begin{align}
b^i &= -\oint du^i(\x) = -\oint dx^k\d_ku^i(\x) \notag\\
&= -\iint dx^j\!\wedge dx^k(\d_j\d_k-\d_k\d_j)u^i(\x),
\end{align}
where $u^i(\x)=x^i-\bar x^i(\x)$ is the displacement vector expressed in terms of displaced coordinates.
When the deformation $\bar x^i(\x)\to x^i$ is regarded as a singular coordinate transformation, an induced vierbein is identified with $e^a{}_i(\x)=\d\bar x^a/\d x^i=\delta^a{}_i-\d_iu^a(\x)$ and leads to the relation of
\begin{align}
b^i = \iint dx^j\!\wedge dx^k
\underbrace{e_a{}^i(\x)[\d_je^a{}_k(\x)-\d_ke^a{}_j(\x)]}_{T^i{}_{jk}(\x)}{} + O(\d u^2)
\end{align}
under Eq.~(\ref{eq:torsion}) for $\omega^{\alpha\beta}{}_\mu(x)=0$.
Therefore, the torsion in the absence of the spin connection is realized as the area density of dislocations at $O(\d u)$.%
\footnote{On the other hand, the Riemann curvature is realized as the area density of disclinations~\cite{Katanaev:2005,Kleinert}.}

In particular, when a single dislocation line exists along the $x^3$ axis, the torsion is provided by
\begin{align}
T^i{}_{12}(\x) = b^i\,\delta(x^1)\delta(x^2),
\end{align}
where $i=1,2$ correspond to the edge dislocation and $i=3$ to the screw dislocation~\cite{Landau-Lifshitz}.
It is only the screw dislocation that can contribute to Eq.~(\ref{eq:axial}) as
\begin{align}\label{eq:potential}
S^0(\x) = -\frac{b^3}{4}\delta(x^1)\delta(x^2), \qquad
S^i(\x) = 0,
\end{align}
which locally serves as an axial chemical potential for Dirac fermions.

\section{Current under magnetic field}\label{sec:magnetic}
\subsection{Derivation}
\begin{figure}[t]
\includegraphics[width=0.9\columnwidth]{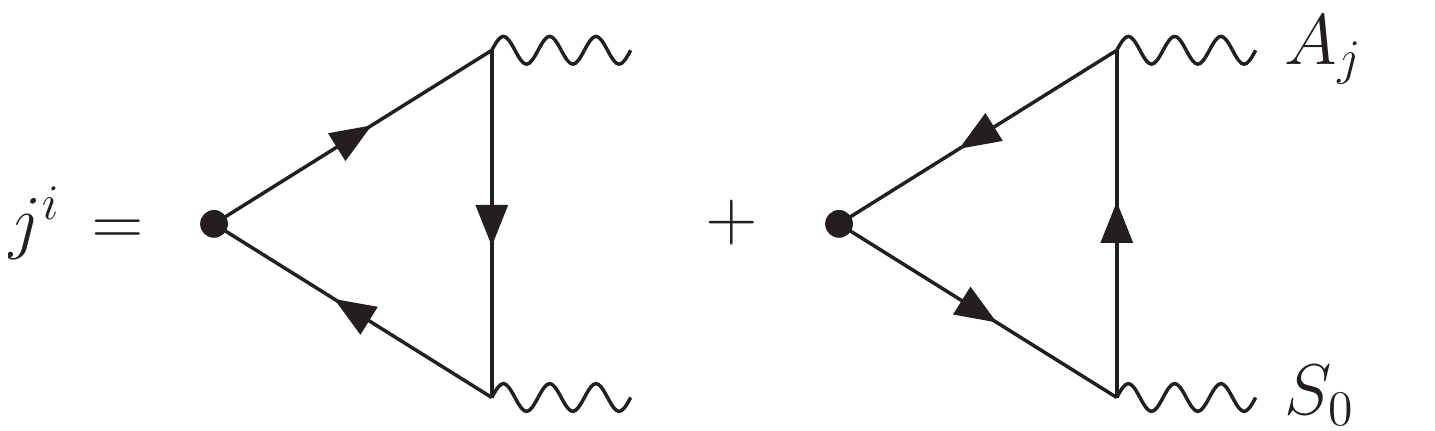}
\caption{\label{fig:current}
Feynman diagrams contributing to the current density in Eq.~(\ref{eq:current}).}
\end{figure}

Motivated by the observations in Sec.~\ref{sec:torsion}, we now study the current density of massless Dirac fermions in flat spacetime with the torsion kept as $S^\mu(x)$ but without $A_5^\mu(x)$, whose action from Eq.~(\ref{eq:action_new}) reads
\begin{align}
S = \int d^4x\,\bar\psi(x)
\bigl[i\slashed\d - \slashed A(x) - \slashed S(x)\gamma^5\bigr]\psi(x).
\end{align}
In particular, we assume $S^\mu(x)=(S^0(\x),\0)$ according to Eq.~(\ref{eq:potential}) and $A^\mu(x)=(0,\A(\x))$ with $B^i=-\epsilon^{0ijk}\d_jA_k(\x)$ being a constant magnetic field and work at the linear order both in $S^0(\x)$ and in $A^i(\x)$.
The current density in equilibrium at finite temperature and chemical potential is obtained from the Feynman diagrams in Fig.~\ref{fig:current} as
\begin{align}\label{eq:current}
j^i(\x) = T\sum_{p_0}\int_{\p,\q,\k}I^{ij0}(p,\q,\k)\,
e^{i\q\cdot\x}\tilde A_j(\q)\,e^{i\k\cdot\x}\tilde S_0(\k).
\end{align}
Here, $p_0=i(2n+1)\pi T+\mu$ with $n\in\Z$ is the fermionic Matsubara frequency, the tildes indicate the spatial Fourier transforms, and the integrand is provided by
\begin{align}
& I^{\mu\nu\kappa}(p,\q,\k) \notag\\
&= \tr\!\left[\frac1{\slashed p+\slashed k-m}\gamma^\kappa\gamma^5
\frac1{\slashed p-m}\gamma^\nu
\frac1{\slashed p-\slashed q-m}\gamma^\mu\right]_\reg \notag\\
&\, + \tr\!\left[\frac1{\slashed p+\slashed q-m}\gamma^\nu
\frac1{\slashed p-m}\gamma^\kappa\gamma^5
\frac1{\slashed p-\slashed k-m}\gamma^\mu\right]_\reg
\end{align}
with $q_0=k_0=0$.

In order to define the superficially divergent integral, we employ the Pauli-Villars regularization, which indicated by the subscript ``reg'' replaces the integrand as
\begin{align}\label{eq:regularization}
\int\,[h(m)]_\reg \equiv \lim_{m\to\infty}\int\,[h(0)-h(m)]
\end{align}
with the limit taken after the integration~\cite{Bertlmann}.
This regularization amounts to subtracting the contribution of massive Pauli-Villars ghosts from that of massless Dirac fermions.
We then expand the integrand with respect to $\q$, leading to
\begin{align}
j^i(\x) = iT\sum_{p_0}\int_{\p,\k}\frac\d{\d q_l}I^{ij0}(p,\q,\k)\bigg|_{\q=\0}
\d_lA_j(\x)\,e^{i\k\cdot\x}\tilde S_0(\k),
\end{align}
where the other terms vanish because of the gauge invariance for the zeroth order and the constant magnetic field for the higher orders.%
\footnote{Here, the regularization is essential to show the gauge-variant zeroth-order term to vanish, which without the regularization turns out to be $j^i(\x)|_{O(A)}=-\epsilon^{0ijk}A_j(\x)\d_kS_0(\x)/(2\pi^2)$.\label{foot:invariance}}
By evaluating the trace as
\begin{align}
& \tr[(\slashed p+\slashed k+m)\gamma^\kappa\gamma^5
(\slashed p+m)\gamma^\nu(\slashed p+m)\gamma^\lambda(\slashed p+m)\gamma^\mu] \notag\\
&= -\tr[(\slashed p+m)\gamma^\lambda(\slashed p+m)\gamma^\nu
(\slashed p+m)\gamma^\kappa\gamma^5(\slashed p+\slashed k+m)\gamma^\mu] \notag\\
&= 4i(p^2-m^2)p_\alpha(p_\beta+k_\beta)
[g^{\kappa\lambda}\epsilon^{\alpha\beta\mu\nu}
+ g^{\kappa\nu}\epsilon^{\alpha\beta\lambda\mu}
+ g^{\lambda\nu}\epsilon^{\alpha\beta\kappa\mu} \notag\\
&\qquad + g^{\alpha\kappa}\epsilon^{\beta\lambda\mu\nu}
- g^{\alpha\lambda}\epsilon^{\beta\kappa\mu\nu}
+ g^{\alpha\nu}\epsilon^{\beta\kappa\lambda\mu}
- g^{\beta\mu}\epsilon^{\alpha\kappa\lambda\nu}] \notag\\
&\quad - 4i(p^2-m^2)m^2\epsilon^{\kappa\lambda\mu\nu}
- 16ip_\alpha k_\beta p^\lambda p^\nu\epsilon^{\alpha\beta\kappa\mu}
\end{align}
and adapting it for $(\mu,\nu,\kappa,\lambda)=(i,j,0,l)$ and $k_0=0$,%
\footnote{In addition to the standard trace technology~\cite{Peskin-Schroeder}, the identify of $g^{\alpha\beta}\epsilon^{\kappa\lambda\mu\nu}+g^{\alpha\kappa}\epsilon^{\lambda\mu\nu\beta}+g^{\alpha\lambda}\epsilon^{\mu\nu\beta\kappa}+g^{\alpha\mu}\epsilon^{\nu\beta\kappa\lambda}+g^{\alpha\nu}\epsilon^{\beta\kappa\lambda\mu}=0$ is applied, which appears referred to as the Schouten identity~\cite{Hattori:2019}.}
we obtain
\begin{align}
j^i(\x) &= -8T\sum_{p_0}\int_{\p,\k}
\left[\frac{(p_0^2-m^2)\delta^i{}_j - (p^i+k^i)p_j}
{(p_0^2-E_{\p+\k}^2)(p_0^2-E_\p^2)^2}\right]_\reg \notag\\
&\quad \times B^je^{i\k\cdot\x}\tilde S_0(\k)
\end{align}
with $E_\p=\sqrt{\p^2+m^2}$.

The Matsubara frequency summation can be performed with the residue theorem after replacing it by the complex contour integration and leads to
\begin{align}
& T\sum_{p_0}\frac1{(p_0^2-E_{\p+\k}^2)(p_0^2-E_\p^2)} \notag\\
&= \frac{N_+(E_{\p+\k})-1}{2E_{\p+\k}(E_{\p+\k}^2-E_\p^2)}
- \frac{N_+(E_\p)-1}{2E_\p(E_{\p+\k}^2-E_\p^2)}
\end{align}
and
\begin{align}
& T\sum_{p_0}\frac1{(p_0^2-E_{\p+\k}^2)(p_0^2-E_\p^2)^2} \notag\\
&= \frac{N_+(E_{\p+\k})-1}{2E_{\p+\k}(E_{\p+\k}^2-E_\p^2)^2}
- \frac{N_+(E_\p)-1}{2E_\p(E_{\p+\k}^2-E_\p^2)^2} \notag\\
&\quad + \frac{N_+(E_\p)-1}{4E_\p^3(E_{\p+\k}^2-E_\p^2)}
- \frac{N'_+(E_\p)}{4E_\p^2(E_{\p+\k}^2-E_\p^2)},
\end{align}
where $N_+(E)\equiv n_T(E-\mu)+n_T(E+\mu)$ is introduced with $n_T(E)=1/(e^{E/T}+1)$ being the Fermi-Dirac distribution function.
By changing the integration variable as $\p\to-\p-\k$ for the terms proportional to $N_+(E_{\p+\k})-1$,%
\footnote{Although this procedure leaves an artificial singularity at $E_{\p+\k}=E_\p$ in Eq.~(\ref{eq:integral}), the resulting integral is correct if it is understood to be the Cauchy principal value, which we confirmed numerically too.
The same remark also applies to Eq.~(\ref{eq:singularity}).\label{foot:singularity}}
we obtain
\begin{align}\label{eq:integral}
j^i(\x) &= 8\int_{\p,\k}
\left[\frac{N_+(E_\p)-1}{2E_\p(E_{\p+\k}^2-E_\p^2)}\delta^i{}_j\right. \notag\\
%&\qquad\left.{} + \frac{N_+(E_\p)-1}{2E_\p(E_{\p+\k}^2-E_\p^2)^2}
%(p^ik_j - k^ip_j)\right. \notag\\
&\qquad\left.{} - \frac{N_+(E_\p)-1}{4E_\p^3(E_{\p+\k}^2-E_\p^2)}
[\p^2\delta^i{}_j - (p^i+k^i)p_j]\right. \notag\\
&\qquad\left.{} + \frac{N'_+(E_\p)}{4E_\p^2(E_{\p+\k}^2-E_\p^2)}
[\p^2\delta^i{}_j - (p^i+k^i)p_j]\right]_\reg \notag\\
&\quad \times B^je^{i\k\cdot\x}\tilde S_0(\k).
\end{align}

The angular integration of $\p$ can be performed by choosing its polar axis to be $\hat\k$ and leads to
\begin{align}\label{eq:angular}
\int\!\frac{d\Omega_\p}{2\pi}\frac1{E_{\p+\k}^2-E_\p^2}
= \frac1{2pk}\ln\left|\frac{2p+k}{2p-k}\right|
\end{align}
and
\begin{align}
& \int\!\frac{d\Omega_\p}{2\pi}
\frac{\p^2\delta^i{}_j - (p^i+k^i)p_j}{E_{\p+\k}^2-E_\p^2} \notag\\
&= \frac{p}{2k}\left[\left(\frac{k}{2p} + \frac{12p^2-k^2}{8p^2}
\ln\left|\frac{2p+k}{2p-k}\right|\right)\delta^i{}_j\right. \notag\\
&\quad\left.{} - \left(\frac{k}{2p} - \frac{4p^2+k^2}{8p^2}
\ln\left|\frac{2p+k}{2p-k}\right|\right)\hat k^i\hat k_j\right]
\end{align}
with $p=|\p|$ and $k=|\k|$.
Furthermore, the radial integration of $\p$ can be performed for the terms without the distribution functions and leads to
\begin{align}\label{eq:radial}
& 8\int_\p\left[-\frac{\delta^i{}_j}{2E_\p(E_{\p+\k}^2-E_\p^2)}
+ \frac{\p^2\delta^i{}_j - (p^i+k^i)p_j}{4E_\p^3(E_{\p+\k}^2-E_\p^2)}\right] \notag\\
&= -\frac{\delta^i{}_j}{4\pi^2}\left[1+\frac{4m^2}{k\sqrt{k^2+4m^2}}
\arctanh\left(\frac{k}{\sqrt{k^2+4m^2}}\right)\right] \notag\\
&\quad + \frac{\hat k^i\hat k_j}{4\pi^2}\left[1-\frac{4m^2}{k\sqrt{k^2+4m^2}}
\arctanh\left(\frac{k}{\sqrt{k^2+4m^2}}\right)\right].
\end{align}
After the integration by parts with respect to $p$, we finally arrive at the current density in the form of
\begin{align}\label{eq:magnetic}
j^i(\x) &= \frac1{4\pi^2}\int_\k\left[1 + \int_0^\infty\!dp\,
N'_+(p)\frac{p}{k}\ln\left|\frac{2p+k}{2p-k}\right|\right] \notag\\
&\quad \times [B^i - \hat k^i(\hat\k\cdot\B)]\,e^{i\k\cdot\x}\tilde S_0(\k),
\end{align}
which constitutes the main outcome of this section.
We note that a similar formula at $\mu=0$ for spacetime-dependent $S_0(x)$ was obtained in Ref.~\cite{Horvath:2020} with the Schwinger-Keldysh formalism.

\subsection{Results}
\begin{figure}[t]
\includegraphics[width=0.9\columnwidth]{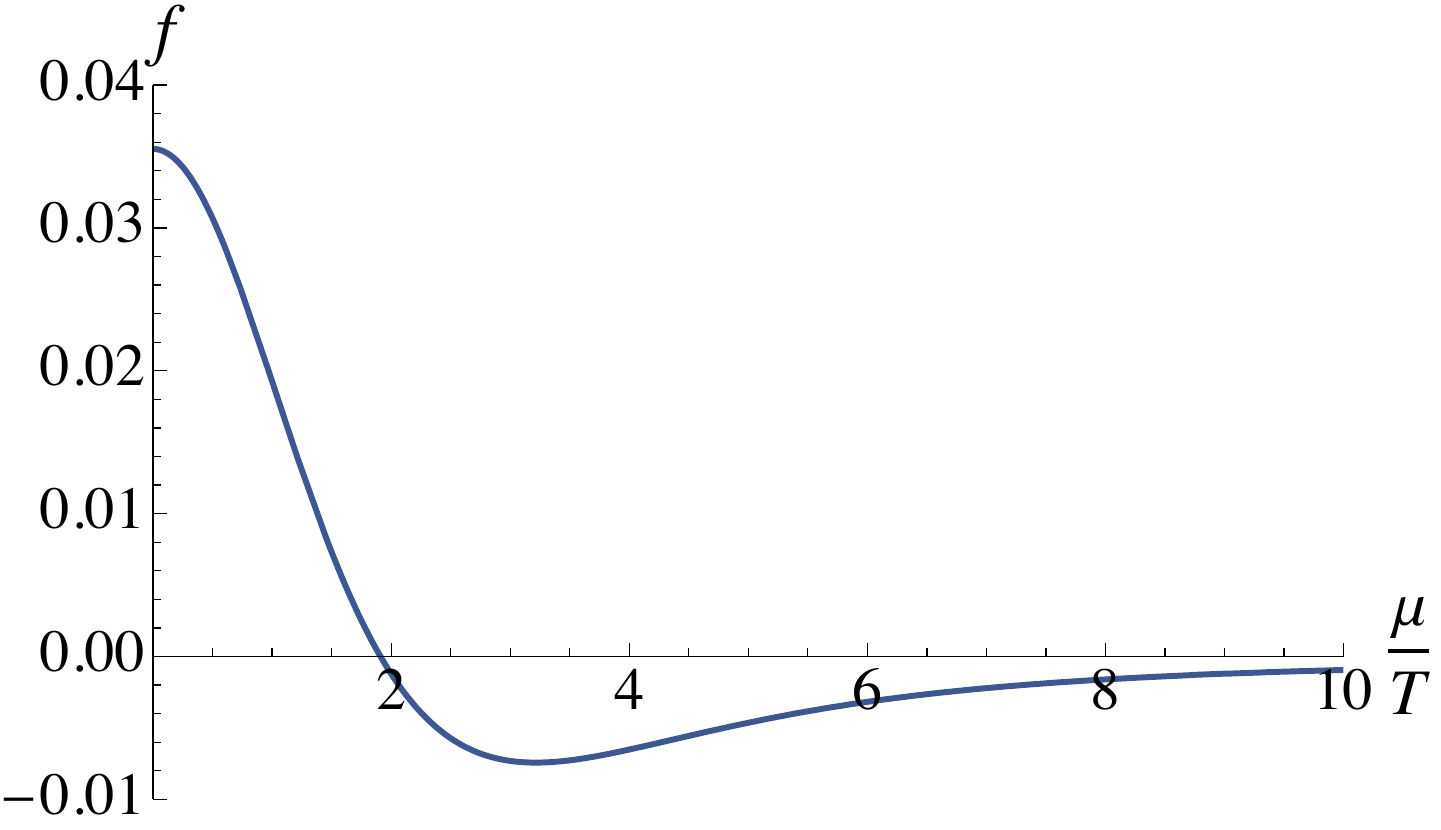}
\caption{\label{fig:expansion}
$f(\mu/T)$ in Eq.~(\ref{eq:expansion}), which reads $f(0)\simeq0.0355$ and $f(\mu/T)\simeq-(T/\mu)^2/12$ at $\mu/T\gg1$.}
\end{figure}

Let us elaborate on Eq.~(\ref{eq:magnetic}).
First of all, the resulting current density satisfies $\grad\cdot\j(\x)=0$, so that the electric charge is locally conserved.
Because such a current density is expressed by rotation of a vector field,%
\footnote{In fact, $B^i-\hat k^i(\hat\k\cdot\B)=-[\hat\k\times(\hat\k\times\hat\B)]^i$.}
the net current flowing through any closed or infinite surface vanishes (see footnote~\ref{foot:limits} regarding the order of limits).
In particular, the expansion of
\begin{align}\label{eq:expansion}
1 + \int_0^\infty\!dp\,N'_+(p)\frac{p}{k}\ln\left|\frac{2p+k}{2p-k}\right|
= f\!\left(\frac\mu{T}\right)\frac{k^2}{T^2} + O(k^4)
\end{align}
with respect to $k$ leads to
\begin{align}
j^i(\x) = f\!\left(\frac\mu{T}\right)
\frac{-B^i\grad^2 + (\B\cdot\grad)\nabla^i}{4\pi^2T^2}S_0(\x) + O(\nabla^4S_0)
\end{align}
with the dimensionless function $f(\mu/T)$ plotted in Fig.~\ref{fig:expansion}.
Accordingly, if a uniform case of $S_0(\x)=-\mu_5$ is considered, the current density itself vanishes, which is consistent with the absence of the chiral magnetic effect in equilibrium~\cite{Vazifeh:2013,Basar:2014,Landsteiner:2016}.

In order to obtain further insight on the last point, we recall that the massive contribution is subtracted from the massless contribution in the integrand of Eq.~(\ref{eq:integral}) according to the regularization in Eq.~(\ref{eq:regularization}).
However, the integral of each contribution over $\p$ is actually convergent as shown in Eq.~(\ref{eq:radial}), so that the current density can be decomposed into
\begin{align}
j^i(\x) = j^i_\D(\x) - j^i_\PV(\x),
\end{align}
where $j^i_\D(\x)$ and $j^i_\PV(\x)$ are contributed by massless Dirac fermions and massive Pauli-Villars ghosts, respectively.
They are obtained by substituting $m=0$ and $m\to\infty$ into Eq.~(\ref{eq:integral}) without the subscript ``reg'', leading to
\begin{align}
j^i_\D(\x) &= -\frac{B^i}{2\pi^2}S_0(\x) + j^i(\x), \\
j^i_\PV(\x) &= -\frac{B^i}{2\pi^2}S_0(\x)
\end{align}
with the latter unaccompanied by the distribution functions because of $\lim_{m\to\infty}N_+(E_\p)=0$.
We now find that $j^i_\D(\x)$ in the uniform case of $S_0(\x)=-\mu_5$ conforms to the chiral magnetic current in Eq.~(\ref{eq:CME}), but is exactly cancelled by $j^i_\PV(\x)$.
Therefore, the regularization is essential for the absence of the chiral magnetic effect in equilibrium as well as for the gauge invariance of the current density (see footnote~\ref{foot:invariance}).
We also note that $j^i_\D(\x)$ and $j^i_\PV(\x)$ are not conserved individually, so that only the sum of them is physically relevant.
In particular, the latter is known as the Bardeen-Zumino or Chern-Simons current, which is independent of the regularization scheme as long as the gauge invariance is respected~\cite{Landsteiner:2016}.

\begin{figure}[t]
~\,\includegraphics[width=0.9\columnwidth]{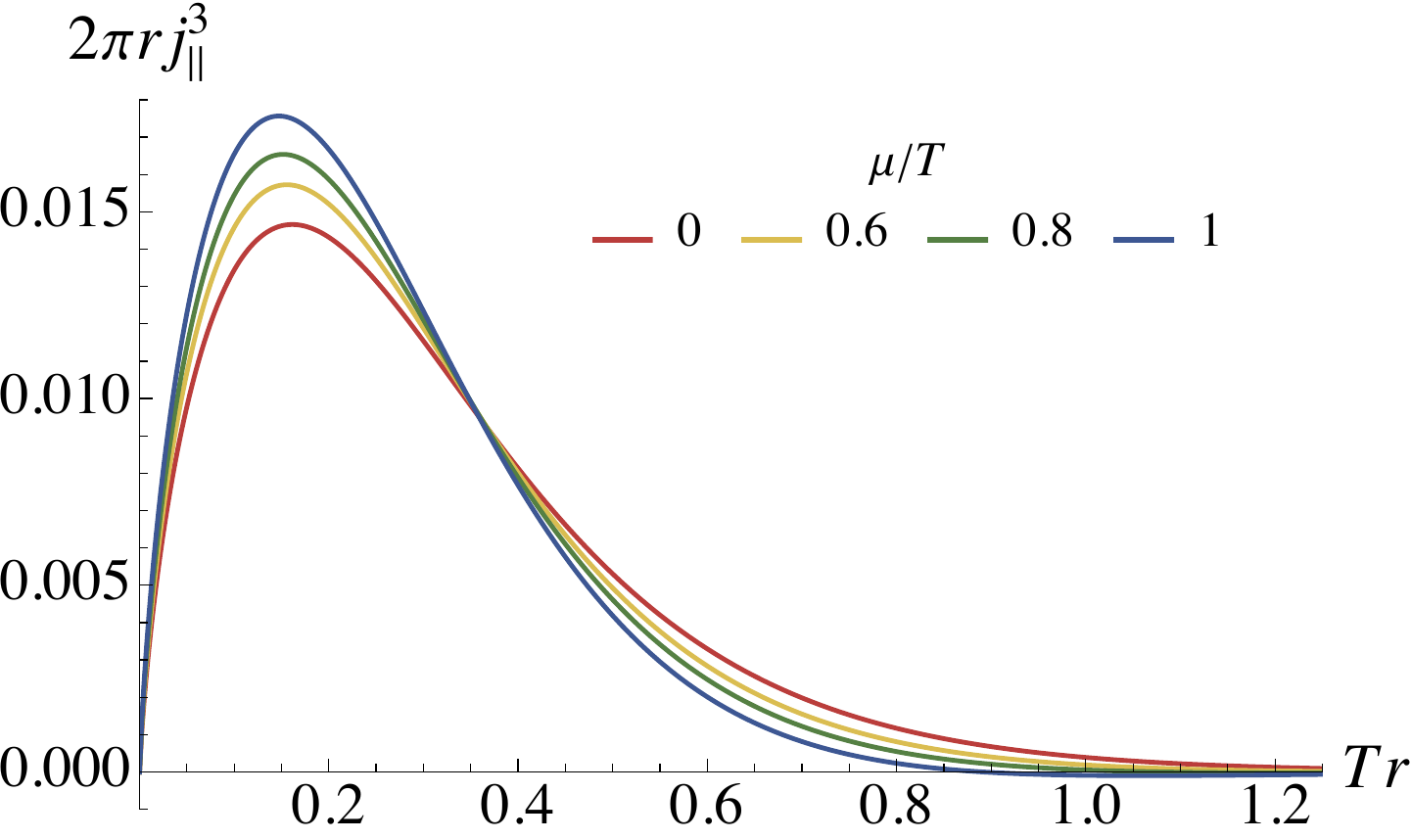}\medskip\\
\includegraphics[width=0.91\columnwidth]{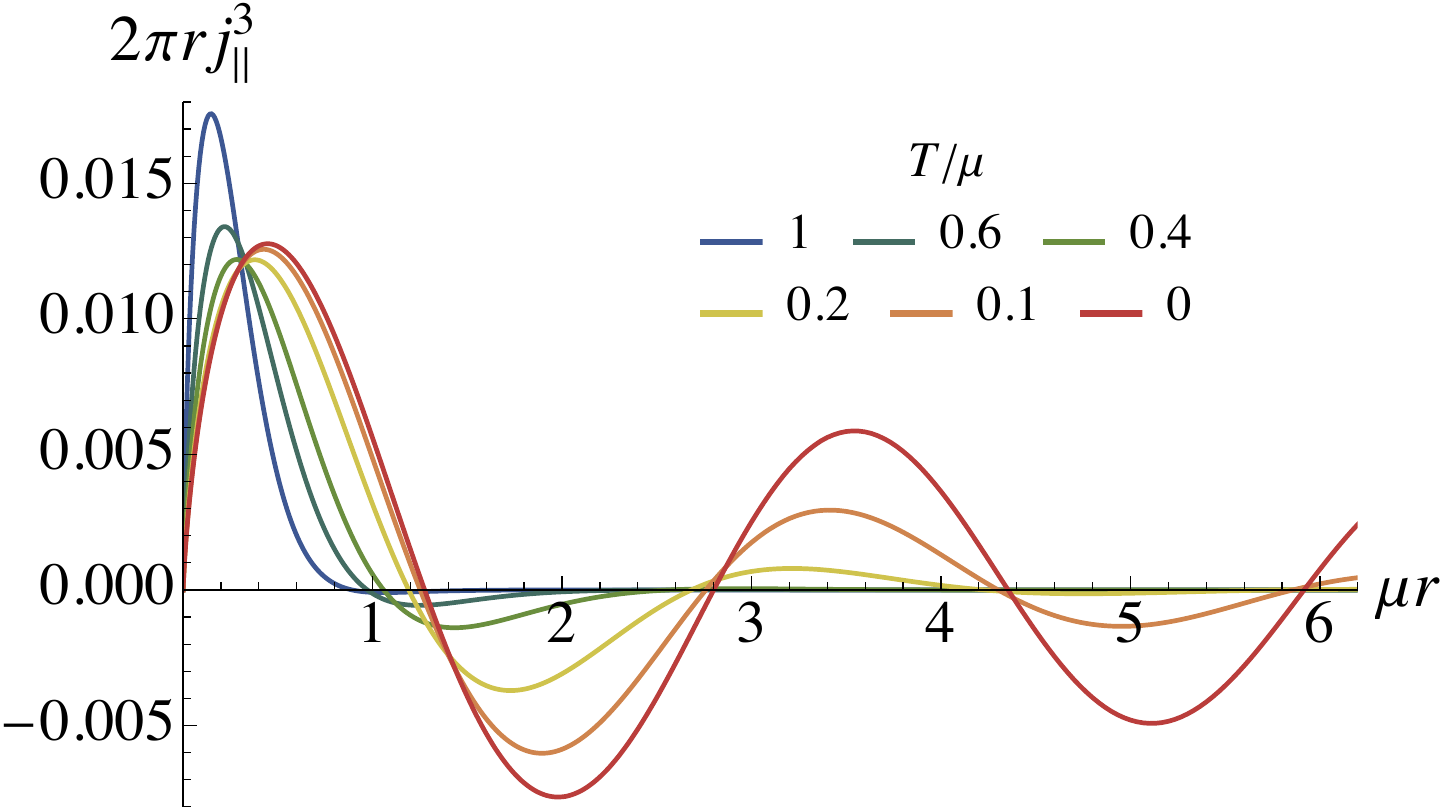}~~
\caption{\label{fig:parallel}
$2\pi rj^3_\parallel(\x)$ at $r\neq0$ resulting from Eq.~(\ref{eq:parallel}) in units of $b^3B^3T$ as a function of $Tr$ for $\mu/T=0$, 0.6, 0.8, 1 (upper panel) and in units of $b^3B^3\mu$ as a function of $\mu r$ for $T/\mu=1$, 0.6, 0.4, 0.2, 0.1, 0 (lower panel).
Its integral $\int_{r\neq0}dr\,2\pi rj^3_\parallel(\x)=b^3B^3/(16\pi^2)$ is independent of $T$ and $\mu$.}
\end{figure}

We then apply Eq.~(\ref{eq:magnetic}) to the case of Eq.~(\ref{eq:potential}), where the axial chemical potential is locally provided by a screw dislocation line along the $x^3$ axis via the torsion.
When the external magnetic field is parallel to the dislocation line, $\B=(0,0,B^3)$, the current density reads $j^1_\parallel(\x)=j^2_\parallel(\x)=0$ and
\begin{align}\label{eq:parallel}
j^3_\parallel(\x) &= -\frac{b^3B^3}{16\pi^2}\delta(x^1)\delta(x^2)
- \frac{b^3B^3}{32\pi^3}\int_0^\infty\!dk\,k\,J_0(kr) \notag\\
&\quad \times \int_0^\infty\!dp\,N'_+(p)\frac{p}{k}\ln\left|\frac{2p+k}{2p-k}\right|
\end{align}
with $r=\sqrt{(x^1)^2+(x^2)^2}$.%
\footnote{Here and below, one should be reminded that $b^3$ and $B^3$ are the third components of the Burgers vector and the magnetic field, respectively, instead of their third powers.}
In addition to the current density localized on the dislocation line, there exists an extended current distribution induced at finite temperature and chemical potential.
The latter for various choices of $\mu/T$ is shown in Fig.~\ref{fig:parallel}, where the Friedel oscillation is found to develop by increasing $\mu/T$.
However, the extended current distribution is absent at zero temperature and chemical potential because of the vanishing distribution function,
\begin{align}\label{eq:vacuum}
j^3_\parallel(\x)|_{T,\mu=0} = -\frac{b^3B^3}{16\pi^2}\delta(x^1)\delta(x^2),
\end{align}
so that only the current density localized on the dislocation line remains in the vacuum.%
\footnote{This apparent contradiction to the vanishing net current arises from the fact that the limits of $T,\mu\to0$ and $R\to\infty$ do not commute for $I=\int_{r<R}dr\,2\pi rj^3_\parallel(\x)$.
Because the extended current density $\sim[\max(T,\mu)]^2$ is distributed over $r\lesssim[\max(T,\mu)]^{-1}$, we obtain $\lim_{R\to\infty}\lim_{T,\mu\to0}I=-b^3B^3/(16\pi^2)$ from Eq.~(\ref{eq:vacuum}) but $\lim_{T,\mu\to0}\lim_{R\to\infty}I=0$ from Eq.~(\ref{eq:parallel}).\label{foot:limits}}

\begin{figure}[t]
\includegraphics[width=0.8\columnwidth]{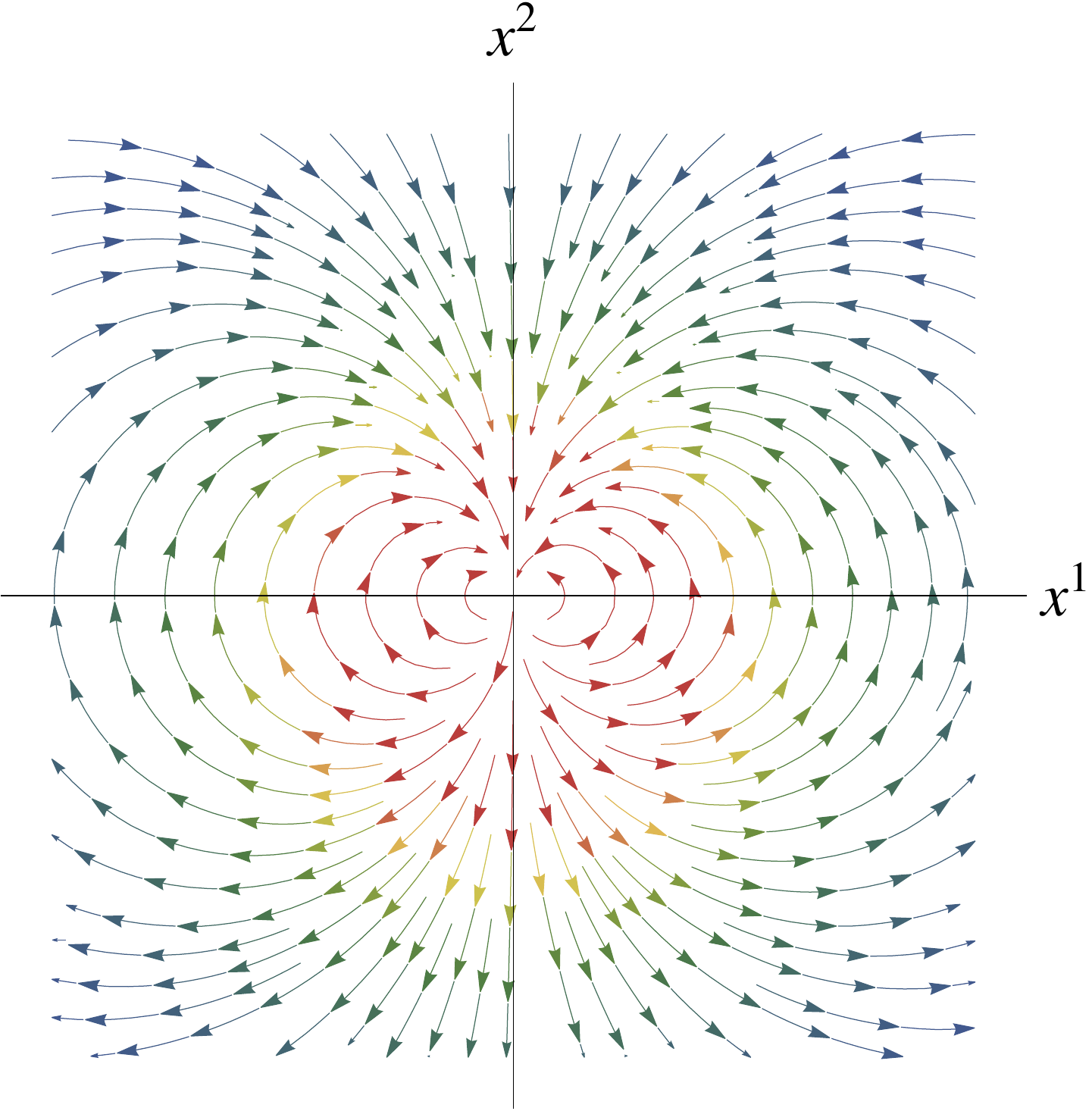}
\caption{\label{fig:perpendicular}
Dipole-like flow of $\j_\perp(\x)|_{T,\mu=0}$ at $r\neq0$ in the $x^1x^2$ plane resulting from Eq.~(\ref{eq:perpendicular}) for the magnetic field applied in the $x^2$ direction.}
\end{figure}

On the other hand, when the external magnetic field is perpendicular to the dislocation line, $\B=(B^1,B^2,0)$, the current density reads $j^3_\perp(\x)=0$ and
\begin{align}\label{eq:perpendicular}
j^i_\perp(\x)|_{T,\mu=0} &= -\frac{b^3B^i}{32\pi^2}\delta(x^1)\delta(x^2)
+ \frac{b^3}{32\pi^3}\frac{B^i - 2\hat x^i(\hat\x\cdot\B)}{r^2}
\end{align}
at zero temperature and chemical potential for $i=1,2$.
In this case, an extended current distribution exists even in the vacuum as shown in Fig.~\ref{fig:perpendicular}.

\section{Current under Weyl node separation}\label{sec:weyl}
We found in Sec.~\ref{sec:magnetic} that the torsion under an external magnetic field locally induces an equilibrium current $\sim B^iS_0(\x)$ akin to the chiral magnetic current.
Here, we explore another possible form of equilibrium current $\sim \mu A_5^iS_0(\x)$, where constant $A_5^i$ amounts to introducing a separation between two Weyl nodes in momentum space~\cite{Miransky:2015}.
Because $B^i$ and $\mu A_5^i$ have the same dimension and the same transformations under charge conjugation, parity, and time reversal~\cite{Peskin-Schroeder}, such a current density is allowed from the symmetry perspective.
In fact, the current density localized around a screw dislocation line was numerically observed in a tight-binding model of Weyl semimetals~\cite{Kodama:2019}.

With this motivation, we now study the current density of massless Dirac fermions in flat spacetime with the torsion kept as $S^\mu(x)$ but without $A^\mu(x)$, whose action from Eq.~(\ref{eq:action_new}) reads
\begin{align}
S = \int d^4x\,\bar\psi(x)
\bigl[i\slashed\d - \slashed A_5\gamma^5 - \slashed S(x)\gamma^5\bigr]\psi(x).
\end{align}
Again, we assume $S^\mu(x)=(S^0(\x),\0)$ according to Eq.~(\ref{eq:potential}) and $A_5^\mu=(0,\A_5)$ being constant and work at the linear order both in $S^0(\x)$ and in $A_5^i$.
The current density in equilibrium at finite temperature and chemical potential is obtained from the Feynman diagrams similar to Fig.~\ref{fig:current} as
\begin{align}
j^i(\x) = T\sum_{p_0}\int_{\p,\k}J^{ij0}(p,\k)\,A_{5j}\,e^{i\k\cdot\x}\tilde S_0(\k).
\end{align}
Here, the integrand is provided by
\begin{align}
& J^{\mu\nu\kappa}(p,\k) \notag\\
&= \tr\!\left[\frac1{\slashed p+\slashed k-m}\gamma^\kappa\gamma^5
\frac1{\slashed p-m}\gamma^\nu\gamma^5
\frac1{\slashed p-m}\gamma^\mu\right]_\reg \notag\\
&\, + \tr\!\left[\frac1{\slashed p-m}\gamma^\nu\gamma^5
\frac1{\slashed p-m}\gamma^\kappa\gamma^5
\frac1{\slashed p-\slashed k-m}\gamma^\mu\right]_\reg
\end{align}
with $k_0=0$ and the Pauli-Villars regularization is employed according to Eq.~(\ref{eq:regularization}).
By evaluating the trace as
\begin{align}
& \tr[(\slashed p+\slashed k+m)\gamma^\kappa\gamma^5
(\slashed p+m)\gamma^\nu\gamma^5(\slashed p+m)\gamma^\mu] \notag\\
&= \tr[(\slashed p+m)\gamma^\nu\gamma^5(\slashed p+m)\gamma^\kappa\gamma^5
(\slashed p+\slashed k+m)\gamma^\mu] \notag\\
&= 8m^2p_\alpha(g^{\alpha\kappa}g^{\mu\nu} - g^{\alpha\mu}g^{\kappa\nu}) \notag\\
&\quad + 4(p^2+m^2)(p_\beta+k_\beta)
(g^{\kappa\mu}g^{\beta\nu} - g^{\kappa\nu}g^{\beta\mu} - g^{\mu\nu}g^{\beta\kappa}) \notag\\
&\quad - 8p_\alpha(p_\beta+k_\beta)p^\nu
(g^{\alpha\beta}g^{\kappa\mu} - g^{\alpha\kappa}g^{\beta\mu} - g^{\alpha\mu}g^{\beta\kappa})
\end{align}
and adapting it for $(\mu,\nu,\kappa)=(i,j,0)$ and $k_0=0$, we obtain
\begin{align}
j^i(\x) &= -8T\sum_{p_0}\int_{\p,\k}
\left[\frac{(p_0^2-E_\p^2)p_0\delta^i{}_j - 2p_0(2p^i+k^i)p_j}
{(p_0^2-E_{\p+\k}^2)(p_0^2-E_\p^2)^2}\right]_\reg \notag\\
&\quad \times A_5^j\,e^{i\k\cdot\x}\tilde S_0(\k).
\end{align}

The Matsubara frequency summation can be performed with the residue theorem after replacing it by the complex contour integration and leads to
\begin{align}
T\sum_{p_0}\frac{p_0}{(p_0^2-E_{\p+\k}^2)(p_0^2-E_\p^2)}
= \frac{N_-(E_{\p+\k})-N_-(E_\p)}{2(E_{\p+\k}^2-E_\p^2)}
\end{align}
and
\begin{align}
& T\sum_{p_0}\frac{p_0}{(p_0^2-E_{\p+\k}^2)(p_0^2-E_\p^2)^2} \notag\\
&= \frac{N_-(E_{\p+\k})-N_-(E_\p)}{2(E_{\p+\k}^2-E_\p^2)^2}
- \frac{N'_-(E_\p)}{4E_\p(E_{\p+\k}^2-E_\p^2)},
\end{align}
where $N_-(E)\equiv n_T(E-\mu)-n_T(E+\mu)$ is introduced.
By changing the integration variable as $\p\to-\p-\k$ for the terms proportional to $N_-(E_{\p+\k})$ (see footnote~\ref{foot:singularity}), we obtain
\begin{align}\label{eq:singularity}
j^i(\x) &= 8\int_{\p,\k}
\left[\frac{N_-(E_\p)}{E_{\p+\k}^2-E_\p^2}\delta^i{}_j\right. \notag\\
&\qquad\left.{} + \frac{N_-(E_\p)}{(E_{\p+\k}^2-E_\p^2)^2}
(2p^i+k^i)k_j\right. \notag\\
&\qquad\left.{} - \frac{N'_-(E_\p)}{2E_\p(E_{\p+\k}^2-E_\p^2)}
(2p^i+k^i)p_j\right]_\reg \notag\\
&\quad \times A_5^j\,e^{i\k\cdot\x}\tilde S_0(\k).
\end{align}

The angular integration of $\p$ can be performed by choosing its polar axis to be $\hat\k$ and leads to Eq.~(\ref{eq:angular}),
\begin{align}
\int\!\frac{d\Omega_\p}{2\pi}\frac{(2p^i+k^i)k_j}{(E_{\p+\k}^2-E_\p^2)^2}
= \frac1{2pk}\ln\left|\frac{2p+k}{2p-k}\right|\hat k^i\hat k_j,
\end{align}
and
\begin{align}
& \int\!\frac{d\Omega_\p}{2\pi}\frac{(2p^i+k^i)p_j}{E_{\p+\k}^2-E_\p^2} \notag\\
&= -\frac{p}{2k}\left(\frac{k}{p} + \frac{4p^2-k^2}{4p^2}
\ln\!\left|\frac{2p+k}{2p-k}\right|\right)(\delta^i{}_j + \hat k^i\hat k_j).
\end{align}
After the integration by parts with respect to $p$, we end up with
\begin{align}\label{eq:weyl}
j^i(\x) = 0
\end{align}
for any $m$ in Eq.~(\ref{eq:singularity}) regardless of the regularization.
This outcome of the vanishing current density in the presence of the torsion and the Weyl node separation is unexpected from the symmetry perspective, which in turn indicates that the nonvanishing current density in the presence of the torsion and the external magnetic field is never trivial but sourced from the axial anomaly.

\section{Summary and discussion}\label{sec:summary}
In summary, we studied equilibrium transport properties of massless Dirac fermions at finite temperature and chemical potential in spacetime accompanied by torsion, which in four dimensions couples with Dirac fermions as an axial gauge field.
In particular, we computed the current density at the linear order in the torsion as well as in an external magnetic field with the Pauli-Villars regularization, finding that an equilibrium current akin to the chiral magnetic current is locally induced according to Eq.~(\ref{eq:magnetic}), which shall be referred to as ``torsional magnetic current.''
Here, the regularization is essential for the current density to be conserved, gauge invariant, and consistent with the absence of the chiral magnetic effect in equilibrium.
Such torsion can be realized in condensed matter systems along a screw dislocation line, around which localized and extended current distributions were predicted as shown in Figs.~\ref{fig:parallel} and \ref{fig:perpendicular} so as to be relevant to Dirac and Weyl semimetals.

Furthermore, we computed the current density at the linear order in the torsion as well as in a Weyl node separation, which in Eq.~(\ref{eq:weyl}) turned out to vanish in spite of being allowed from the symmetry perspective.
This unexpected outcome indicates that the nonvanishing torsional magnetic current is never trivial but sourced from the axial anomaly.
On the other hand, we failed to understand the numerical observation of Ref.~\cite{Kodama:2019}, i.e., the current density localized around a screw dislocation line in a tight-binding model of Weyl semimetals, from the effective field theory perspective, which remains to be resolved in future work.

Finally, we note that various forms of torsion-induced transport have been explored in previous work~\cite{Hughes:2011,Kimura:2012,Hughes:2013,Parrikar:2014,Sumiyoshi:2016,Khaidukov:2018,Ferreiros:2019,Ishihara:2019,Huang:2019,Imaki:2019,Nissinen:2019,Nissinen:2020a,Nissinen:2020b,Huang:2020a,Huang:2020b,Laurila:2020,Imaki:2020,Huang:preprint,Manes:2021,Ferreiros:2021,Chernodub:2022,Nissinen:2022}, some of which on electric current can be contrasted with our findings.
First, the current density called the torsional chiral magnetic effect,
\begin{align}\label{eq:TCME}
j^i_\mathrm{TCME} = \frac\Lambda{4\pi^2}A_{5a}\epsilon^{ijk}T^a{}_{jk},
\end{align}
was proposed with $\Lambda$ being a momentum cutoff~\cite{Sumiyoshi:2016,Ishihara:2019,Huang:2019}.
However, we do not corroborate the existence of Eq.~(\ref{eq:TCME}) because the torsion must appear in the form of $S^\kappa\sim\epsilon^{\kappa\lambda\mu\nu}T_{\lambda\mu\nu}$ as shown in Sec.~\ref{sec:torsion}.
Second, the current density called the chiral torsional effect,
\begin{align}\label{eq:CTE}
j^i_\mathrm{CTE} = \frac2{\pi^2}\mu\mu_5S^i,
\end{align}
was proposed besides a term $\sim T^\nu{}_{\mu\nu}$ unconformable to the allowed form of $S^\mu$~\cite{Huang:2019,Imaki:2019}.
This current density is actually equivalent to $j^i\sim\mu S_0A_5^i$ because the torsion serves as an axial gauge field, which however vanishes as shown in Sec.~\ref{sec:weyl}.
Therefore, we do not corroborate the existence of Eq.~(\ref{eq:CTE}), which is also consistent with the recent claim of Refs.~\cite{Ferreiros:2021,Chernodub:2022}.
Third, the current density called the torsional magnetic effect,
\begin{align}\label{eq:TME}
j^i_\mathrm{TME} = \frac{S_0}{2\pi^2}B^i,
\end{align}
was proposed as an immediate analog of the chiral magnetic effect in Eq.~(\ref{eq:CME})~\cite{Imaki:2020}.
Again, we do not corroborate the existence of Eq.~(\ref{eq:TME}) because it contradicts our Eq.~(\ref{eq:magnetic}) as well as the absence of the chiral magnetic effect in equilibrium~\cite{Vazifeh:2013,Basar:2014,Landsteiner:2016}.
Hopefully, our firm analyses help to clarify possible forms of equilibrium current induced by the torsion in the system of massless Dirac fermions.

\acknowledgments
This work was supported by JSPS KAKENHI Grants No.\ JP18H05405 and No.\ JP21K03384, as well as by Advanced Research Center for Quantum Physics and Nanoscience, Tokyo Institute of Technology.

\end{document}